\documentclass{mn2e}
\usepackage{psfig}
\newcommand\beq{\begin{equation}}
\newcommand\eeq{\end{equation}}
\newcommand\kpc{{\rm \,kpc}}

\newcommand\Mpc{$h^{-1}$\,{\rm \,Mpc}}
\newcommand\Msun{$h^{-1}\,M_\odot$}
\newcommand\mnras{MNRAS}
\newcommand\apj{ApJ}
\newcommand\apjs{ApJS}

\title[The formation of dark matter haloes]{Formation time distribution of dark matter haloes: theories versus N-body simulations}

\author[W.P. Lin, Y.P. Jing \& Lihwai Lin]{
W.P.~Lin,$^1$$^2$\thanks{{\it E-mail} address: linwp@center.shao.ac.cn, ypjing@center.shao.ac.cn, d90222005@ms90.ntu.edu.tw}
   Y.P.~Jing$^1$$^2$, Lihwai~ Lin$^3$\\
   $^1$ Shanghai Astronomical Observatory, the Partner Group of Max-Planck-Institute f\"ur Astrophysik, \\ Nandan Road 80, Shanghai 200030, China \\
   $^2$ Department of Astronomy \& Beijing Astrophysics Center, Peking University, Beijing, 100871, China \\
   $^3$ Department of Physics, National Taiwan University, Taipei
}
\date{}
\pagerange{\pageref{firstpage}--\pageref{lastpage}}
\pubyear{2003}

\begin{document}
\maketitle
\label{firstpage}

\begin{abstract}
This paper uses numerical simulations to test the formation time
distribution of dark matter haloes predicted by the analytic
excursion set approaches. The formation time distribution is
closely linked to the conditional mass function and this test is
therefore an indirect probe of this distribution. The excursion set
models tested are the extended Press-Schechter (EPS) model, the
ellipsoidal collapse (EC) model, and the non-spherical collapse
boundary (NCB) model. Three sets of simulations (6 realizations)
have been used to investigate the halo formation time distribution
for halo masses ranging from dwarf-galaxy like haloes ($M=10^{-3} M_*$, 
where $M_*$ is the characteristic non-linear mass
scale) to massive haloes of $M=8.7 M_*$. None of the models
can match the simulation results at both high and low redshift. In
particular, dark matter haloes formed generally earlier in our
simulations than predicted by the EPS model. This discrepancy
might help explain why semi-analytic models of galaxy formation,
based on EPS merger trees, under-predict the number of high
redshift galaxies compared with recent observations.
\end{abstract}


\section{Introduction}
\label{introduction}
In the present preferred cosmological model, dark matter
haloes form hierarchically through accretion and merging of smaller
structures that grow from a Gaussian initial density field.  This
process is modeled by Press-Schechter theory\cite{PS}, which
simply assumes a region with the mass over-density above a certain
threshold will turn around and eventually collapse to form a bound
object. Bond et al.(1991) developed an excursion set approach
and used it to derive the number density of collapsed dark matter
haloes more rigorously. Lacey \& Cole (1993) used the excursion set
theory to predict the merger rate at which small objects merge with
each other to form larger objects, the conditional probability of
its progenitor for a parent halo, as well as the survival probability of
haloes. The predictions for the halo formation time distribution
have been tested against N-body simulations [with $128^3$ particles]
\cite{LC94}.  The conditional probability predicted by the excursion
set theory (or EPS theory) has been widely used to plant merger
trees \cite{KW93,LC94,SK99} so as to construct semi-analytical models
of galaxy formation, and to study the clustering of dark matter halos
(Mo \& White 1996; Mo, Jing, \& White 1997).  
However the EPS theory may fail to model the details of halo formation.
For instance, the predicted mass function was not found to fit the
simulation results very well (e.g. Lee \& Shandarin 1998, Sheth \&
Tormen 1999). Tormen (1998) also reported that the excursion set
predictions did not well fit the conditional mass function of
sub-clumps in simulations.  In addition, since the EPS theory failed
to correctly describe the spatial distribution of small haloes in high
resolution numerical simulations (Jing 1998, 1999; Lee \& Shandarin
1999; Sheth \& Tormen 1999), it may also fail to predict the halo
formation time distribution.  There are already some indications that the
distribution of halo formation time in N-body simulations is not
consistent with predictions\cite{vdB02a}.

In fact, it is an approximation that the spherical collapse assumed in the EPS formalism.
This process can be better modeled by a triaxial turn-around model \cite{BM96,SMT01,ST02}. 
One of such models is the so-called Ellipsoidal Collapse model or EC model.
By taking the ellipsoidal collapse into account, Sheth, Mo and Tormen (2001) found that the
modified mass function can fit N-body simulations well.  
Sheth \& Tormen (2002) pointed out that the conditional mass function was not
universal since it was not consistent with their simulation results at every redshift.
Another model is the Non-spherical Collapse Boundary model (or NCB model shortly) 
proposed by Chiueh \& Lee (2001). 
It relates the halo formation to the collapse of the Zel'dovich pancakes. 
A recent version of the model was presented by Lin, Chiueh \& Lee (2002).
Both the EC and the NCB models were calibrated to fit the spatial
two-point correlation function of halos (Jing 1998) and the mass
function (Sheth \& Tormen 1999) over a large range of halo mass.

This paper uses three sets of high resolution N-body simulations [with $256^3$ or $512^3$ particles] to study the distribution of halo formation redshift and make comparison with theory predictions.
It is organized as follows.  In section \ref{theory}, we
present the theory predictions by three analytical models. 
The simulations are described briefly in section \ref{simulations} together with the method to find the halo formation redshift. 
There the simulation results are compared to the predictions.  
Main conclusions and discussion are
given in section \ref{discussion}.

\section{The theoretical predictions for the distribution of halo formation redshift}
\label{theory}

The halo formation redshift is defined as the redshift at which its main progenitor 
has accumulated half of the halo mass. According to the EPS theory,
the probability that a volume of mass $M_1$, which is within the
region of mass $M_2$ collapsed at redshift $z_2$, collapsed to form a
progenitor at redshift $z_1$ is given by the conditional probability
function,

\[f_{S_1}(S_1,\delta_{c1}|S_2,\delta_{c2})
dS_1=\frac{(\delta_{c1}-\delta_{c2})}{\sqrt{2\pi}(S_1-S_2)^{3/2}}\times
\]
\beq
\exp\left[-\frac{(\delta_{c1}-\delta_{c2})^2}{2(S_1-S_2)}\right]dS_1
\eeq
\[
(S_1 >S_2, \delta_{c1}> \delta_{c2})
\]
where $\delta_{c1}\equiv \delta_c(z_1)$ and $\delta_{c2}\equiv
\delta_c(z_2)$ \cite{LC93}.  Here $\delta_{c}(z)$ is
the critical over-density required for the spherical collapse at redshift
$z$ (see eq.A6), and $S_i \equiv \sigma^2(M_i)$ where
$\sigma(M_i)$ is the rms of the initial density fluctuation field
smoothed on a scale which contains mass $M_i$, extrapolated using
linear theory to the present time (see eq. A1).

Thus the conditional probability can be converted into the
probability that a halo of mass $M_2$ at redshift $z_2$ has a
progenitor in the mass range $M_1$ to $M_1 +dM_1$ at an earlier epoch
$z_1$ (i.e. the conditional mass function), 
\beq \frac{dp}{dM_1}(M_1,z_1|M_2,z_2)dM_1
=\left(\frac{M_2}{M_1}\right)f_{S_1}(S_1,\delta_{c1}|S_2,\delta_{c2})dS_1\,.
\eeq

When using the excursion set model with ellipsoidal collapse
\cite{ST02}, the conditional probability should be replaced by following function,

\[
f_{S_1}(S_1,z_1|S_2,z_2) dS_1=\frac{|T(S_1,z_1|S_2,z_2)|}{\sqrt{2\pi}(S_1-S_2)^{3/2}}\times
\]
\beq
\exp\left\{-\frac{[B(S_1,z_1)-B(S_2,z_2)]^2}{2(S_1-S_2)}\right\} d S_1,
\eeq
and
\[
T(S_1,z_1|S_2,z_2) = \sum_{n=0}^5 {(S_2-S_1)^n\over n!}
{\partial^n [B(S_1,z_1)-B(S_2,z_2)]\over\partial {S_1}^n}
\]
where the moving barrier $B(S,z)=\sqrt{a S_*}[1+\beta (S/a S_*)^\alpha]$ with $S_*\equiv \delta^2_{c}(z)$.
The parameters are adopted from the best fitting of the mass function with N-body simulations, $a=0.707, \alpha=0.485, \beta=0.615$ \cite{SMT01,ST02}.

While for the NCB model \cite{CL01}, the conditional probability is a fitting formula \cite{LCL02} which reads as,
\beq
f_{S_1}(S_{1},\delta_{c}(z_{1})|S_{2},\delta_{c}(z_{2})) =
f(\mu')\left|\frac{\partial\mu'}{\partial S_{1}}\right|.
\eeq
Here
\beq
\label{fitting}
\mu^{'}f(\mu^{'})d\mu'=2A(\kappa)(1+\frac{1}{\mu^{'^{2q}}
})(\frac{\mu^{'^{2}}}{2\pi})^{1/2}\mathrm{exp}(-\frac{\mu^{'^{2}}}{2})d\mu',
\eeq

where
\beq
\label{mu}
\mu^{'}\equiv\frac{[\delta_{c}(z_{1})-\delta_{c}(z_{2})]\varepsilon(S_{2},\kappa)}{(S_{1}-S_{2})^{1/2}},
\eeq
\beq
\label{eqA}
A=0.322+\frac{0.178}{\kappa},
\eeq
\beq
\label{eqb}
\varepsilon(S_{2},\kappa)=\varepsilon(x)=0.036x^{4}-0.309x^{3}+0.944x^{2}-1.060x+1
\eeq with $x\equiv(\sqrt{S_{2}}/\delta_{c}(z_{2})-0.25)/\kappa$, and
$q(\kappa)$ is required to satisfy the normalization condition 
\beq
A=\sqrt{\frac{\pi}{2}}/(\sqrt{\frac{\pi}{2}}+\frac{\Gamma[-q+\frac{1}{2}]}{2\times\frac{1}{2}
^{(-q+\frac{1}{2})}}).  
\eeq 
Here $\kappa$ is the separation of two boundaries, which is defined as
$\delta_c(z_1)/\delta_c(z_2)$. Refer to the appendix of Lin et
al. (2002) for further explanation of the fitting procedure for these parameters.
Note that we use the uncorrected $\mu^{'}$
[eq.(29) of Lin et al.] rather than the corrected $\mu^{'}$ [eq.(33)
of Lin et al.].  There are two reasons for this choice.
First, the halo formation distribution is incorrectly normalized when using the corrected $\mu'$. Second, the correction for $\mu'$ could be mass dependent.

Integrating equation (2) over the mass range $M_2/2<M_1<M_2$ gives the
probability $P(<t_1)$ that its formation time is earlier than $t_1$ or its
formation redshift is larger than $z_1$.  For a halo with mass $M_0$
formed at the present, we set $M_2=M_0, t_2=t_0, z_2=z_0=0$. Therefore
the probability can be written as
\[
P(<t_f)\equiv P(>z_f)
\]
\beq =\int^{S_h}_{S_0}\frac{M_0}{M(S_1)}f_{S_1}(S_1|S_0)dS_1, 
\eeq
with $S_h=S(M_0/2)$ \cite{LC93}. In this integration, the conditional probability
functions (1), (4) and (5) are used for the EPS, EC and NCB models
respectively.  The accumulative probability $P(>z_f)$ and thereby its redshift distribution $\frac{dP}{dz_f}$, can be calculated numerically.

\begin{figure}
\centerline{\psfig{file=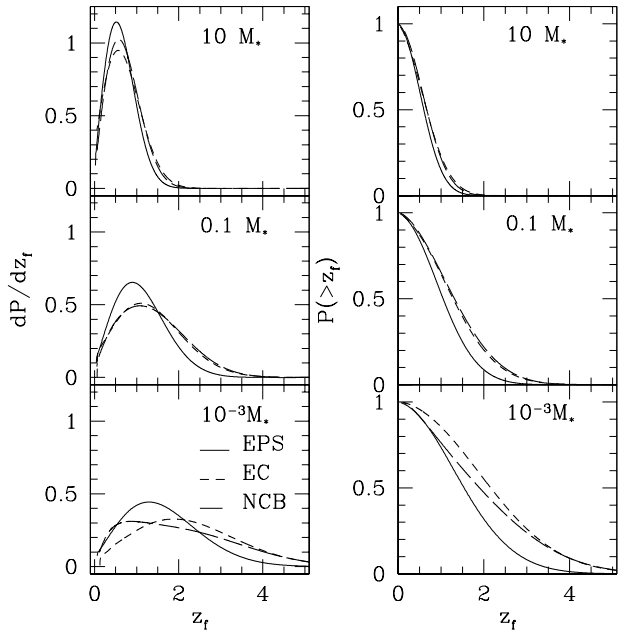,width=9.0cm}}
\caption{ The model predictions of the formation redshift for haloes
with mass of $10^{-3}$, $0.1$ and $10 {M_\star}$. The left panels give
the differential probability distribution of the halo formation
redshift, and the right panels give the accumulative probability of
haloes formed at redshift higher than $z_f$.  }
\label{fig1}
\end{figure}

The predictions of the halo formation redshift distribution for 3 typical masses are shown in Fig.\ref{fig1}. 
Here we assume a LCDM cosmogony (with $\sigma_8=0.9$) whose parameters will be given in next section.  
The left and right panels show respectively the formation redshift distribution and the probability that a halo formed at redshift larger than $z_f$.  
The solid, dotted and dashed lines represent the EPS, EC and NCB predictions respectively.  
As can be seen, for haloes with a small mass of $10^{-3}M_*$, the predictions of
the three models differ dramatically from each other. 
At low redshift the NCB prediction is close to the EPS one, however the predictions by the NCB and EC model are coincident at high redshift.
For haloes with masses of $0.1M_*$ and $10 M_*$, the EC and NCB results only have a small difference.
For the large mass, there is almost no big difference in the
prediction among the three models.  
In general, the $dP/dz_f$ profiles of the EC and NCB models are broader and have lower peaks than those of the EPS model.  
Compared to the EPS model, the EC or the NCB model predicts a larger fraction of haloes formed at high redshift and the EC model predicts a smaller fraction at low redshift. 

\section{Simulations}
\label{simulations}

Three samples of N-body simulations are used to study the formation
redshifts of dark haloes with mass ranging from $10^{-3}\,M_*$ to
$8.7\, M_*$.  Each simulation has at least 30 outputs, and can be used
to trace the formation of a halo accurately.  The cosmological model
is the currently popular flat low-density model with the density
parameter $\Omega_0=0.3$ and the cosmological constant $\Lambda_0=0.7$
(LCDM).  The shape parameter of the linear density power spectrum is
$\Gamma \equiv \Omega_0 h=0.2$.
The characteristic mass $M_*$ is $9.55\times 10^{12}${\Msun} for
LCDMa, LCDMc simulations and $1.66\times 10^{13}${\Msun} for LCDMb
simulations.  Other parameters of the simulations are listed in Table
\ref{A}, where $\sigma_8$ is the amplitude of the power spectrum, 
$N$ is the total number of particles in the simulation
boxes, $m_p$ is the mass of a particle and $z_i$ is the initial
redshift of the simulations.

The simulation data were generated on the VPP5000 Fujitsu
supercomputer of the national Astronomical Observatory of Japan with a
vectorized-parallel P$^3$M code (Jing \& Suto 1998, Jing \& Suto 2002).  

\begin{table*}
\caption{Model parameters for simulations}
\label{A}
\begin{tabular}{ccccccccc}
\hline 
Model &$\sigma_8$ & N & box-size & $m_p$ & time-steps& $z_i$ & outputs & realizations \\
 & & &  \Mpc& & \Msun & & &\\
\hline
LCDMa &0.9 & $256^3$ & 25 & $7.73\times 10^7$ & 5000 & 72 &169 & 2\\
LCDMb &1.0 & $256^3$ & 100& $4.95\times 10^9$ &  600 & 36 & 30 & 3\\
LCDMc &0.9 & $512^3$ & 300& $1.67\times 10^{10}$ & 1200 & 36 & 60 & 1\\
\hline
\end{tabular}
\end{table*}

\subsection{The formation of small dark matter haloes}
\label{simulation1}
We use the LCDMa simulations to study the formation of small dark
haloes.  For each simulation, 56 out of the 169 outputs are used.
Only those haloes with 100 particles or more at the present are
included to assure a reliable identification of haloes (i.e., at the
half-mass formation redshift, they have 50 particles at least). The
haloes with more than 1200 particles will not be considered, since these
haloes are not abundant enough to have a reliable determination of the
distribution of the halo formation redshift.  Thus the dark haloes we
study here have a mass range between $7.73\times 10^9$ and $9.28 \times
10^{10}$ {\Msun} ($\sim 10^{-3} -10^{-2} M_*$).  These haloes are
``small'' and represent the typical dark haloes of dwarf galaxies.

Dark matter haloes are identified with the spherical over-density
method \cite{JS02}. We adopt the following method to find the
formation redshift of a halo.  At the beginning we pick up a halo as a
parent halo in the final output ($z=0$), and find out the particles
within its virial radius.  Then the member particles are traced in the
outputs from high redshift to low redshift step by step.  We calculate
the fraction of the member particles in all progenitors, and select
the progenitor with the maximum number of particles as the main
progenitor.  Generally, the members of the main progenitor will
increase with time because of merging and accretion, although in a few
circumstances its mass may decrease because of mass loss due to the
tidal stripping by nearby haloes and/or unbound particles. When the
main progenitor has reached at least half of the parent halo's mass
for the first time, we define the corresponding redshift as the
formation redshift (half-mass formation redshift $z_f$) of the
selected parent. Alternatively, we can go along the merger tree from
redshift $0$ to high redshift, and we can define the formation
redshift as the time when the mass of the main progenitor first drops
below half of the parent halo's mass. We find that these two
definitions give almost an identical distribution of the halo
formation time distribution\footnote{There is only a slightly shift of the
profile, but the conclusions change little.}, as serious tidal
stripping of the main progenitor happens only rarely. In the
following, we will adopt the first definition and compare the
simulations with the three analytical models.

\begin{figure}
\centerline{\psfig{file=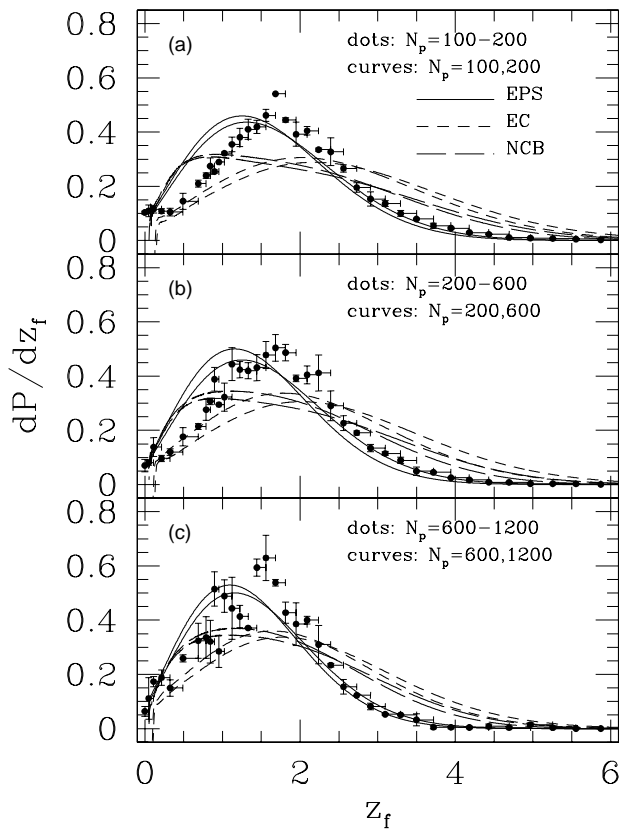,width=9.0cm}}
\caption{The distribution of the halo formation redshift for dwarf-galaxy
like haloes. The points with error bars are the simulation
results, and the lines are the predictions of the three
analytical models. Different panels show the results for haloes of
$N_p$ particles.  The total number of simulated haloes (2
realizations) used for the analysis are 2334/2295, 1747/1786, 505/456
for the results in panel (a)-(c) respectively.}
\label{fig2}
\end{figure}

The results are shown in Fig. \ref{fig2}(a)-(c).  The solid points are the
result measured from the simulations, and the vertical error bars
present the $1\sigma$ error of the mean value derived from the scatter
between the realizations.  The solid, short-dashed, and long-dashed
lines are the predictions of the EPS, EC and NCB analytical
models. For the model predictions, we plotted two lines for halo at the
lower mass end (the curve with higher peak) and the higher mass end (the
curve with lower peak) respectively.  We plot both of them, because
there may be a bias when using the prediction for the mean halo mass,
as the formation probability is invariant in terms of
$\sigma(M)^2$(rather than $M$).  For the narrow ranges of halo mass
chosen in this study, this bias is sufficiently small as the
predictions at the upper and lower mass limits are very close.  To
elucidate another possible bias caused by the limited output
resolution, we also add a right-hand ``error'' bar for the simulation
results.  This bias is introduced by the fact the halo formation time
in simulation is measured to be the redshift at which the fraction of
the member particles is larger than 0.5 for the first time, however
the real formation redshift should fall between this output and the
earlier one.  This uncertainty is shown by the right-hand ``error''
bars.

As the figure shows, at low redshift ($\le 1$) the EPS and NCB models predict more halos than the simulations, while the EC predictions are consistent with the simulation results. However at high redshift, the EPS model fits the simulation results better than the EC and NCB models.
In every panel, the formation redshift distribution of the simulations
peaks at a higher redshift than the EPS prediction
\footnote{For the theoretical predictions of the halo formation
redshift, a test on the infrared cut-off of wavenumber $k$ due to a
finite simulation box has been done.  The cut-off can modify the $M
-\sigma^2(M)$ relation so that it can change the distribution of the
halo formation redshift.  However, for the small haloes considered
here, the wavenumber cutoff has negligible effect on their formation
redshift.} and has a narrower profile (but with a higher peak) than the EC and NCB predictions.

\subsection{The formation of sub-$M_*$ dark matter haloes}
\label{simulation2}

\begin{figure}
\centerline{\psfig{file=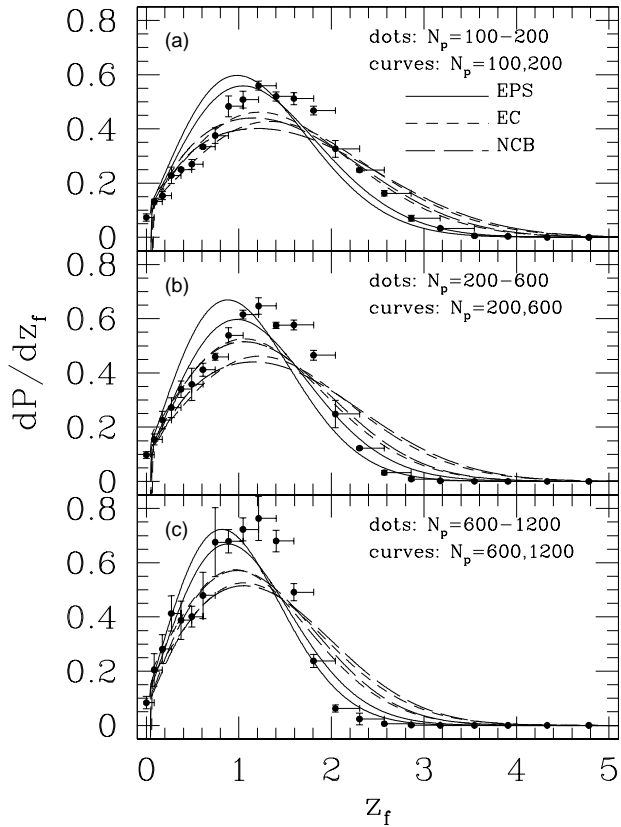,width=9.0cm}}
\caption{The same as Fig.\ref{fig2}, but for sub-$M_*$ haloes.
The total number of simulated haloes (3 realizations) used for the
analysis are 3577/3487/3607, 2698/2549/2619, 758/773/724 for the
results in panel (a)-(c) respectively.  }
\label{fig3}
\end{figure}

The LCDMb simulations are used to study the formation of sub-$M_*$
dark haloes.  Again the number of particles in a halo spans from 100
to 1200, or the corresponding halo mass ranges from $4.95\times
10^{11}$ {\Msun} to $5.94\times 10^{12}$ {\Msun} (i.e. $\sim 0.03-0.36
M_*$).  The definition of haloes is slightly different from that used
in the last subsection.  Here distinct haloes were found using the FOF
method with a bonding length 0.2 times of the mean particle
separation. We have tested the results for one realization against the
two halo identification algorithms, and found that the two
identifications give nearly identical results. We calculate the halo
formation redshift as in the last subsection for the 30 outputs, and
plot the results in Fig.\ref{fig3}.

The results found for the sub-$M_*$ halos are quite similar to those
for smaller haloes, but we can see continuous changes with
halo mass of the model predictions relative to the simulation results. The
EPS model predicts too many haloes of low formation redshifts again,
while the EC and NCB models predict relatively well the fraction of
these haloes. On the other hand, the EC and NCB models predict too many
haloes of high formation redshift.  
Note that for the mass range chosen here the EC and NCB predictions are close to each other
(cf. Fig.\ref{fig1}).

\subsection{The formation of large dark matter haloes}
\label{simulation3}

\begin{figure}
\centerline{\psfig{file=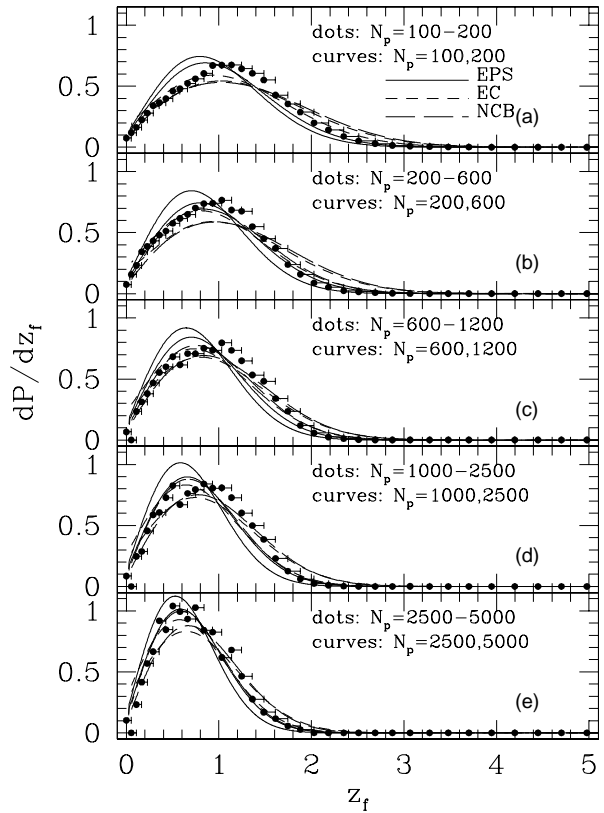,width=9.0cm}}
\caption{The same as Fig.\ref{fig2}, but with $512^3$ particles in the
simulation box and for haloes with mass $\ge 0.17 M_*$.  The total
number of simulated haloes used for the analysis are 35381, 25418,
7060, 5334, and 1840 in panels (a)-(e) respectively.  }
\label{fig4}
\end{figure}

We use a simulation (LCDMc) with $512^3$ particles (Jing 2002) to study the
formation of large dark haloes.
The corresponding halo mass ranges from $1.67\times 10^{12}$ {\Msun}
to $8.35\times 10^{13}$ {\Msun} ($\sim 0.17-8.74 M_*$). We did the
same analysis as the last subsection, and plot the results in
Fig.\ref{fig4}.  Since the LCDMc simulation has only one realization,
no error bars are plotted for $dP/dz_f$.  However, because of the high
mass resolution, the population of haloes within each mass range is so
large that the random fluctuation among the bins is small.  This also
makes us be able to extend the analysis to massive haloes with up to
5000 member particles.

Figure \ref{fig4} continuously shows the change of the formation redshift
distribution with the halo mass.  The pattern of the differences
between simulation results and predictions is similar to that of small dark haloes.
For the very massive haloes [Fig.\ref{fig4}(e)], the differences among
the three models are small, and they agree with the simulation results
reasonably well.  The results shown in Fig. \ref{fig3}(c) has mass
range of $0.18-0.36 M_*$ which is almost the same as that of
Fig. \ref{fig4}(a) (whose mass range is $0.17-0.35M_*$), even though
$\sigma_8$ used in these simulations are slightly
different. Therefore, these two results can be compared for the same
population of haloes with different mass resolutions.  As can be seen,
the results are consistent within the errorbars, though the population
in the LCDMb simulation appears to form slightly later that that in
the LCDMc.

\subsection{Resolution tests}

\begin{figure}
\centerline{\psfig{file=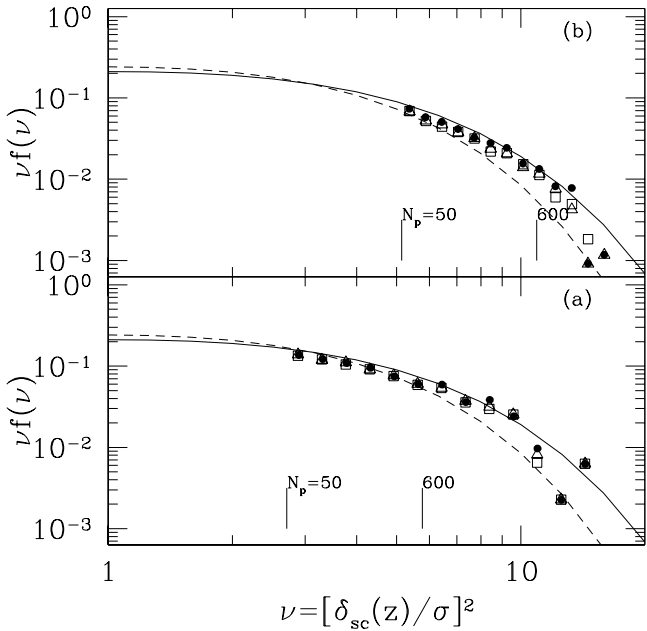,width=9.0cm}}
\caption{(a) The mass function of haloes at $z\sim 3.18$ in the three
simulations: the dots for the simulation with $z_i =72$, the triangles
for the simulation with the force softening $78$ {\kpc}, and the open
squares for the first LCDMb realization.  The lines labeled with $N_p$
indicate the $\nu$ values corresponding to 50 and 600 member particles
respectively. The dashed line and solid line represent the predictions
by the EPS model and the EC model.  (b) The same as (a), but at $z\sim
4.78$.}
\label{fig5}
\end{figure}

We consider halos with at least 100 particles, so the identification
of the halos is generally secure. There are still a couple of issues
one should consider about the simulation resolutions.  If the
simulations are not started sufficiently early, some non-linear
collapse could be missed or delayed, and the initial distribution of
particles might still have an appreciable effect at the high redshift.
Another worry is that some low mass haloes may be missed because of
the force softening adopted in the simulations.  To check these
issues, we performed another two simulations that were run until $z
\simeq 3.18$. These two simulations have the same model parameters and
the same initial fluctuations(including the phase) as the first
realization of the LCDMb simulations, but one simulation was started
at an earlier epoch $z_i=72$ and the other adopts force softening
$\eta=78\kpc$ that is twice large of the value used in the LCDMb
simulation ($39 \kpc$). The two simulations have also the data output
at $z \simeq 4.78$.  In Figure \ref{fig5}, we plot the mass functions $\nu
f(\nu)$ of haloes at both redshifts for these two simulations, and
compared them with that of the corresponding LCDMb realization, where
$\nu \equiv [\delta_c(z)/\sigma(m)]^2$. Except for very massive haloes
that are rare objects (so their mass functions have large statistical
fluctuations), the agreement of the mass functions among the three
simulations is nearly perfect, especially for the mass range of 50 to
600 member particles that we are interested in this paper.  Therefore
we conclude that our results here are robust to the starting redshifts
and to the force softening of the simulations.

\section{discussion and summary}
\label{discussion}

This work was initially motivated by the fact that the semi-analytical models of galaxy formation which were constructed based on the EPS theory for the merger tree
predict systematically redder colours for dwarf galaxies than the observations 
of faint blue galaxies (van den Bosch 2002b).  
It is known that the EPS theory fails accurately describing the mass
function and the spatial correlation of small haloes ($M\ll M_*$, Jing
1998, Lee \& Shandarin 1998, Sheth \& Tormen 1999). 
This difficulty motivated many theorists to re-formulate the formation of haloes, and reproduce the mass function as well as the spatial two-point correlation function of haloes in numerical simulations (Sheth et al. 2001; Chiueh \& Lee 2001, Lin et al. 2002).
It is therefore interesting to test if the EPS theory, or the alternative
models can correctly predict the distribution of the halo formation time 
(which is more closely related to the problem of faint blue galaxies).

With the help of a large set of high-resolution N-body simulations, we
have found various degrees of success and failure for the analytical
models at different halo mass.
This is not unexpected since the excursion set approaches 
roughly model the growth of haloes but overlook some factors which could be important.  
For example, the models overlook the correlation between the fluctuations on different scales, do not include tidal striping, etc. 
If the conditional mass function has problems, the merger trees constructed upon it may not 
correctly describe the growth of haloes (especially for small haloes) and
therefore should be used with caution in galaxy formation models.  
We suggest that future studies of the conditional mass function should be
pushed toward lower mass end using simulations with high mass
resolution (like those used here).

The implications of our results for the problem of the faint blue
galaxies are quite clear. If the merger trees used by the
semi-analytical models are replaced by the merger trees from the
simulations, one would expect that the dwarf galaxies will become
generally redder, which contradicts with the observations more
seriously. There are several ways to solve this discrepancy. First,
the observed faint blue galaxies may not constitute a representative
sample of small haloes. This could happen if many of red dwarf
galaxies, because of their low star formation rate, have escaped from
being detected. Second, star formation within small haloes is
significantly delayed due to heating or re-ionization (see Mo\& Mao 2002, and reference therein).
Third, faint blue galaxies may experience a recent interaction with nearby galaxies, 
which will trigger star formation. 
We have traced the trajectories of small haloes, and found
that about 10\% $\sim$ 15\% of the small haloes once pass the
central part (with distance less than half of a virial
radius) of a bigger halo which is at least 3 times more massive than the
small one.  
These haloes should have experienced a strong interaction.
If gas in the small halo would not be stripped off, these strong interactions could trigger star formation and change the colour of the dwarf galaxies to blue.
However these processes are so complex that more efforts are needed to work out the
problem of dwarf galaxies.

From the results of the LCDMb simulations, we found an earlier formation time for sub-$M_*$
haloes (about $10^{12}$ solar mass) than the EPS prediction. 
This may indicate that galaxies are formed slightly earlier
in the numerical simulations than in the EPS model. 
This result can explain a recent finding of Cimatti et al. (2002).
Cimatti et al. measured the redshift distribution of galaxies with $K<20$, and
compared it with the predictions of semi-analytical models of galaxy
formation based on the EPS merger trees. They found that the galaxies
in the observation formed earlier than in the galaxy formation models
of about 0.3 in redshift. Qualitatively this can be seen in our Figure 3.  
Of course, the formation times of haloes and galaxies are not the same
thing, but they should have a similar trend.

There seems to be discrepancy between our results and those of van den
Bosch (2002a) who found that the formation time distribution of the haloes in the
GIF simulation deviates from the EPS prediction more severely at a
larger mass.  In fact, he checked for the haloes in two mass ranges:
$5.6\times 10^{11}\le M\le 1.1\times 10^{12}$\Msun and $2.0\times
10^{12}\le M\le 2.0\times 10^{14}$\Msun (his Figure 4), corresponding
to the number of particles $40\le N_p\le 80$ and $140\le N_p\le 14000$
respectively.  The number of haloes in each mass bin is dominated by
the lower mass end, especially in the larger mass bin.  For the small
haloes of $M\le 1.1\times 10^{12}$\Msun, it is likely that his
formation time is underestimated (i.e. formation delayed) because of
the limited resolution, and so his agreement with the EPS theory
becomes better. His result for large haloes should be compared with
the middle panel of our Figure 3 (according to the effective
mass). This comparison clearly shows that his results agree well with
ours.

In summary, none of the predictions by the EPS, EC and NCB  model can match the simulation results at both low and high redshift.  
This discrepancy implies that the theoretical mass conditional function needs further improvement.  
Our results may have an important impact on the models of galaxy formation.  
The ``blue-color'' problem of dwarf galaxies can hardly be solved solely
by the gravitational interaction between dark matter.
For sub-$M_*$ haloes, an increase of formation
redshift by $\sim 0.3$ relative to the EPS model may relieve the
difficulty of the semi-analytical models to explain the recent observation of K20 galaxies
(Cimatti et al. 2002) at high redshift.

\section*{Acknowledgments}
The authors would like to thank Gerhard B\"orner, Houjun Mo, Tzihong
Chiueh and Jounghun Lee for useful discussions, and the anonymous
referee for helpful comments and suggestions. YPJ is supported by NSFC
No.10125314 and NKBRSF G19990754. WPL acknowledges the grants provided
by NKBRSF G1999075402, NFSC No.10203004 and Shanghai NFS
No. 02ZA14093.

{\appendix
\section{The linear growth of the density perturbation}

\bigskip
We assume that the universe is dominated by cold dark matter. 
The mass $M$ of a halo is related with the co-moving length
scale $r_0$ using a top-hat filter in real space $M=(4\pi/3)\pi
r_0^3\rho_0$, where $\rho_0$ is the co-moving mean mass density of the
Universe.  The variance of the density fluctuations, smoothed over a
region of mass M, is given by
\beq
  \sigma^2(r_0) = \int_0^\infty \frac{dk}{k} \,  k^3 P(k) W^2(k r_0)
\eeq
where $W(kr_0)$ is the window function for top-hat filtering
\beq
  W(kr_0) = 3 \left[ \frac{\sin(kr_0)}{(kr_0)^3} - \frac{\cos(kr_0)}{(kr_0)^2}
\right] \; .
\eeq
The power spectrum at present is given by
\beq
  P(k) \propto k T^2(k)
\eeq
which uses the self-similar primordial density spectrum with index
$n=1$. For the transfer function, we have used
\[
 T(k) =\frac{\ln (1+2.34q)}{2.34q}\times
\]
\beq
\left[1+3.89q+(16.1q)^2+(5.46q)^3+(6.71q)^4\right]^{-1/4}
\eeq
and
\beq
 q \equiv \frac{k}{\Gamma h^{-1} {\rm Mpc}}
\eeq
\cite{Bardeen86}. 
In practice $\sigma^2(r_0)$ was calculated up to an overall constant which is
fixed by the choice of $\sigma_8 \equiv \sigma(8 h^{-1} {\rm Mpc})$.  

Collapsed halos are taken to be regions in the {\it linear\/} density
field with the density contrast greater than some critical density
contrast, $\delta_c$.  In practice we take into account of the linear
growth by holding the variance $\sigma$ fixed and increasing the
density thresholds at high redshift,
\beq 
\delta_c(t(z)) = (1+z) \frac{g(\Omega_0)}{g(\Omega(z))}
\delta_c \; .  
\eeq

The growth rate can be approximated as
\beq
g(\Omega (z)) =
\frac{5}{2} \Omega \left[ \frac{1}{70} + \frac{209}{140} \Omega -
 \frac{\Omega^2}{140} + \Omega^{4/7} \right]^{-1}
\eeq
and
\beq
\Omega(z) = \Omega_m \frac{(1+z)^3}{1 -\Omega_m + (1+z)^3 \Omega_m}
\eeq
for a flat $\Lambda$ universe \cite{CPT92}.

}


\begin{thebibliography}{99}

\bibitem[Bardeen et al. 1986]{Bardeen86} Bardeen J., Bond J.R., Kaiser N., Szalay A.S., 1986, \apj, 304, 15

\bibitem[Bond \& Myers 1996]{BM96} Bond J.R., \& Myers S., 1996, \apjs, 103, 1

\bibitem[Bond et al. 1991]{Bond91} Bond J.R., Cole S., Efstathiou G., Kaiser N., 1991, \apj, 379, 440

\bibitem[Carroll, Press \& Turner 1992]{CPT92} Carroll S.M., Press W.H., \& Turner E.L., 1992, ARA\&A, 30, 499

\bibitem[Chiueh \& Lee 2001]{CL01}Chiueh T., \& Lee J., 2001, \apj, 555, 83

\bibitem[Cimatti, et al. 2002]{CI02} Cimatti A., et al., 2002, A\&A Letter, 391, L1

\bibitem[Jing 1998]{Jing98} Jing Y.P., 1998, \apj, 503, L9

\bibitem[Jing 1999]{Jing99} Jing Y.P., 1999, \apj, 515, L45
\bibitem[Jing(2002)]{2002MNRAS.335L..89J} Jing, Y.~P.\ 2002, \mnras, 335, 
L89 

\bibitem[Jing \& Suto 1998]{JS98} Jing Y.P., \& Suto Y., 1998, \apj, 494, L5

\bibitem[Jing \& Suto 2002]{JS02} Jing Y.P., \& Suto Y., 2002, \apj, 574, 538

\bibitem[Kauffmann \& White 1993]{KW93} Kauffmann G., \& White S.D.M., 1993, \mnras, 261, 921

\bibitem[Lacey \& Cole 1993]{LC93}Lacey C., \& Cole S., 1993, \mnras, 262, 627

\bibitem[Lacey \& Cole 1994]{LC94}Lacey C., \& Cole S., 1994, \mnras, 271, 676

\bibitem[Lee \& Shandarin 1998]{Lee98} Lee J., \& Shandarin S.F., 1998, \apj, 500, 14

\bibitem[Lee \& Shandarin 1999]{Lee99} Lee J., \& Shandarin S.F., 1999, \apj, 517, 5L

\bibitem[Lin et al. 2002]{LCL02}Lin L., Chiueh T., \& Lee J., 2002, \apj, 574, 527

\bibitem[Mo \& Mao 2002]{MM02} Mo H.J., \& Mao S., 2002, \mnras, 333, 768
\bibitem[Mo \& White(1996)]{1996MNRAS.282..347M} Mo, H.~J.~\& White, 
S.~D.~M.\ 1996, \mnras, 282, 347 
\bibitem[Mo, Jing, \& White(1997)]{1997MNRAS.284..189M} Mo, H.~J., Jing, 
Y.~P., \& White, S.~D.~M.\ 1997, \mnras, 284, 189 


\bibitem[Press \& Schechter 1974]{PS} Press W., \& Schechter P., 1974, \apj, 187, 425

\bibitem[Sheth, Mo \& Tormen 2001]{SMT01} Sheth R.K., Mo H.J., \& Tormen G., 2001, \mnras, 323, 1

\bibitem[Sheth, \& Tormen 1999]{ST99} Sheth R.K., \& Tormen G., 1999, \mnras, 308, 119

\bibitem[Sheth \& Tormen 2002]{ST02} Sheth R.K., \& Tormen G., 2002, \mnras, 329, 64

\bibitem[Somerville \& Kolatt 1999]{SK99} Somerville R.S., \& Kolatt T.S., 1999, \mnras, 305, 1

\bibitem[Tormen 1998]{Tormen98} Tormen G., 1998, \mnras, 297, 648

\bibitem[van den Bosch 2002a]{vdB02a} van den Bosch F.C., 2002a, \mnras, 331, 98

\bibitem[van den Bosch 2002b]{vdB02b} van den Bosch F.C., 2002b, \mnras, 332, 456

\end{thebibliography}
\end{document}